# P-V-T equation of state of epsilon iron and its densities at core conditions

József Garai

*Department of Earth Sciences, Florida International University, Miami, FL 33199, USA*

**Abstract**

The author's new p-V-T Equation of State is tested against the available experiments of epsilon iron. The root-mean-square-deviations (RMSD) of the molar volume, pressure, and temperature are 0.021 cm$^3$, 2.0 GPa and 144.9 K respectively.

Separating the experiments into 200 K ranges the new EoS was compared to the most widely used finite strain, interatomic potential, and empirical isothermal EoSs such as the Burch-Murnaghan, the Vinet, and the Roy-Roy respectively. Correlation coefficients, RMSDs of the residuals and Akaike Information Criteria were used for evaluating the fittings. Based on these fitting parameters the new p-V-T EoS is equal or better than the conventional isothermal EoSs.

The newly defined parameters were used to calculate the density of the inner core. The calculated densities are significantly lower than the current consensus indicating that it might be too early excluding the possibility of a pure iron-nickel core with no light elements.

## 1. Introduction

It has been suggested that the volume is not a fundamental variable but rather the sum of the initial (zero pressure-temperature) $[V_o]$, thermal $[V_o^{th}]$ and elastic $[V_o^{el}]$ volumes (Garai, 2007).

$$V = V_o + V_o^{el} + V_o^{th}. \tag{1}$$

Subscript o is used to indicate that these fundamental volume parts were determined by using the newly defined volume coefficient of thermal expansion $\alpha_o$ and bulk modulus $B_o$.

The thermodynamic equations describing these fundamental volume parts are

$$V_o = nV_o^m \tag{2}$$

and

$$\alpha_o \equiv \frac{1}{V_{p=0}} \frac{\partial V^{th}}{\partial T}, \tag{3}$$

and

$$B_o \equiv -V_{T=0} \frac{\partial p}{\partial V^{el}}, \tag{4}$$

where n is the number of moles, $V_o^m$ is the molar volume of the substance at zero pressure and temperature, T is the temperature, and p is the pressure. The thermal volume can be calculated from Eq. (3) as:

$$V_o^{th} = V_o \left( e^{\int_{T=0}^{T} \alpha_o dT} - 1 \right) \tag{5}$$

while the elastic volume from Eq. (4) as:

$$V_o^{el} = V_o \left( e^{-\int_{p=0}^{p} \frac{1}{B_o} dp} - 1 \right). \qquad (6)$$

Substituting these fundamental volume components into Eq. (1) gives the actual volume [V] as:

$$V = V_o \left( e^{-\int_{p=0}^{p} \frac{1}{B_o} dp} + e^{\int_{T=0}^{T} \alpha_o dT} - 1 \right). \qquad (7)$$

Based on theoretical considerations it has been suggested that the temperature derivatives of the volume coefficient of thermal expansion and the bulk modulus are zero (Garai, 2007). Assuming that the pressure derivatives are constant and introducing the multipliers [a, c] Eq. (7) can be written as:

$$V = V_o \left( e^{-\int_{p=0}^{p} \frac{1}{ap+B_o} dp} + e^{\int_{T=0}^{T} (cp+\alpha_o) dT} - 1 \right). \quad \text{New (5)} \qquad (8)$$

The New (5) EoS was tested to the available experiments of perovskite with positive result. The number in the parenthesis refers to the number of parameters in the equation.

**2. Experiments of epsilon iron**

Iron, the prime constituent of the core, changes its phase and the different polymorphs are stable only within limited pressure and temperature intervals. At core pressures epsilon and possibly beta phase are stable (Nguyen, 2004). Experiments are available in the epsilon phase (Dubrovinsky et al., 1997; 1998; 1998; 2000; Saxena and Dubrovinsky 2000; Saxena et al., 1995; Saxena, 2003, Mao et al. 1987, Mao et al. 1990, N. Funamori et al. 1996; T. Uchida et al.

2001). The 289 experiments span the pressure and temperature range of 11-306 GPa and 293-1573 K respectively. The distribution of the experiments is shown on Fig. 1.

The fitting parameters for the New (5) EoS [Eq. (8)] were calculated. The correlation coefficient is relatively low R = 0.997035 and the RMSD value for the molar volume is 0.059 cm$^3$. The volume coefficient of thermal expansion has negative value which seems to be unreasonable (Table 1). Using hit or miss method many modified version of Eq. (8) were tested. The best fit occurred when an exponential factor [b] has been introduced.

$$V = V_o \left( e^{-\int_{p=0}^{p} \frac{1}{ap^b + B_o} dp + \int_{T=0}^{T} (cp + \alpha_o) dT} - 1 \right). \qquad (9)$$

or after the integration

$$V = V_o \left( e^{\frac{-p}{ap^b + B_o}} + e^{(cp + \alpha_o)T} - 1 \right). \quad \text{New (6)} \qquad (10)$$

The introduction of exponential factor for the volume coefficient of thermal expansion was also tested but it did not improve the fitting. The parameters providing the best fit for New (6) are: $K_o = 116.31$ GPa, $V_o = 6.818$ cm$^3$, $\alpha_o = 2.823 \times 10^{-4}$ K$^{-1}$, a = 6.439, b = 0.7563 and $c = -7.119 \times 10^{-8}$ K$^{-1}$GPa$^{-1}$. The root-mean-square-deviations (RMSD) of the residuals are 0.021 cm$^3$, 2.0 GPa, and 145 K for the molar volume, pressure, and temperature respectively. Based on visual inspections clustering of the experiments conducted by the different labs is noticeable indicating that systematic errors might be present in the experiments (Fig. 2.).

Equation (10) predicts that the volume coefficient of expansion is diminished above a certain pressure that I will call solid critical pressure [$p_{\alpha_o = 0}$].

$$p_{\alpha_o=0} = \left|\frac{\alpha_o}{b}\right| \qquad (11)$$

The diminishing of the volume coefficient of thermal expansion indicates that at pressures higher than the solid critical pressure the interatomic potential energy becomes symmetrical and the temperature does not affect the volume. In order to avoid negative value for the volume coefficient of thermal expansion the introduction of a multiplier [I] is necessary

$$V = V_o \left( e^{\frac{-p}{ap+B_o}} + e^{I(bp+\alpha_o)T} - 1 \right). \quad \text{New (6)} \qquad (12)$$

The multiplier is defined as:

$$\begin{aligned} I &\equiv 1 \quad \text{if} \quad p_{\alpha_o=0} > p \\ &\quad \text{and} \\ I &\equiv 0 \quad \text{if} \quad p_{\alpha_o=0} \leq p \end{aligned} \qquad (13)$$

The New (6) EoS allows calculating any of the variables. The analytical solution of Equation (12) for the temperature is

$$T = I \frac{\ln\left(\frac{V}{V_o} + 1 - e^{\frac{-p}{ap+K_o}}\right)}{bp + \alpha_o}. \qquad (14)$$

The pressure can be determined by repeated substitution until convergence is reached as:

$$p = \lim_{n \to 10} p_{n+1} = -(ap_n + K_o)\ln\left[\frac{V}{V_o} + 1 - e^{I(bp_n+\alpha_o)T}\right] \qquad (15)$$

where

$$n \in \mathbb{N}^* \quad \text{and} \quad p_1 = 0. \qquad (16)$$

The convergence of equation (15) is the function of the pressure. For the maximum pressure used in this study n = 20 gives sufficiently good result.

## 3. Comparing the new EoS to isothermal EoSs

The data set was separated into groups covering 200 K temperature range. Within this temperature range it was assumed that the condition is isothermal. For each temperature range the fitting parameters of the most widely used, finite strain, interatomic potential, and empirical isothermal EoSs [Eqs. (17)-(20)] were determined. Good summary of the contemporary EoSs can be found in Garai (2007).

The most popular finite strain EoS is the Burch-Murnaghan (Murnaghan, 1937, 1944; Birch, 1947)

$$p = \frac{3B_0}{2}\left[\left(\frac{V_0}{V}\right)^{\frac{7}{3}} - \left(\frac{V_0}{V}\right)^{\frac{5}{3}}\right]\left\{1 + \frac{3}{4}(B_0' - 4)\left[\left(\frac{V_0}{V}\right)^{\frac{2}{3}} - 1\right]\right\}. \qquad (17)$$

The universal EoS derived by Rose from a general inter-atomic potential (Rose et al. 1984) and promoted by Vinet (Vinet, 1987a, 1987b) is the most widely used interatomic potential EoS which gives very good results at high pressures.

$$p = 3K_0 \frac{1-f_V}{f_V^2} e^{\left[\frac{3}{2}(B'-1)(1-f_V)\right]} \qquad (18)$$

where

$$f_V = \left(\frac{V}{V_0}\right)^{\frac{1}{3}}. \qquad (19)$$

Roy and Roy (1999) introduced an empirical EoS

$$V = V_0\left[1 - \frac{\ln(1+ap)}{b+cp}\right], \qquad (20)$$

where

$$a = \frac{1}{8B_0}\left[3(B_0'+1)+(25B_0'^2+18B_0'-32B_0B_0''-7)^{\frac{1}{2}}\right]$$

$$b = \frac{1}{8}\left[3(B_0'+1)+(25B_0'^2+18B_0'-32B_0B_0''-7)^{\frac{1}{2}}\right]$$

$$c = \frac{1}{16}\left[3(B_0'+1)+(25B_0'^2+18B_0'-32B_0B_0''-7)^{\frac{1}{2}}\right](B_0'+1)-\frac{1}{8}\left[3(B_0'+1)+(25B_0'^2+18B_0'-32B_0B_0''-7)^{\frac{1}{2}}\right]$$

and demonstrated its applicability to the wide variety of substances. The symbols, $V_0$ and $K_0$ are the volume and the bulk modulus at zero pressure respectively and $K_0'$ is the first derivative of the bulk modulus.

Using the parameters of the New (6) EoS determined from the overall fit, the volume and the pressure was calculated for experiments falling into the 200 K region. From the residuals of the volume and the pressure the RMSD and AIC values were calculated [New (6-0)]. The second number indicates the number of parameters allowed to change for the isothermal fitting. Using fixed values (determined from the overall fitting) for the initial volume, for the volume coefficient of thermal expansion, and the pressure derivative of the volume coefficient of thermal expansion, the fitting parameters of the New (6-3) EoS were calculated.

*3.1. Fitting criteria*

The fitting accuracy of empirical EoSs with the same number of parameters is evaluated by correlation coefficients and/or root-mean-square deviations (RMSDs). The fit quality of models using different numbers of parameters can not be evaluated by their correlation coefficients only (Lindsey, 2004; Burnham and Anderson, 2002, 2004). The test devised assessing the right level of complexity is the Akaike Information Criteria AIC (Akaike, 1973, 1974). Assuming normally distributed errors, the criterion is calculated as:

$$\text{AIC} = 2k + n \ln\left(\frac{\text{RSS}}{n}\right), \tag{21}$$

where n is the number of observations, RSS is the residual sum of squares, and k is the number of parameters. The preferred model is the one which has the lowest AIC value.

Based on the AIC values both New (6-0) and New (6-3) fit better to experiments than the Roy-Roy EoS in every temperature range. The New (6-0) gives better fit two times but weaker fit four times in comparison to the Birch-Murnaghan and Vinet EoSs. The New (6-3) is superior to the Birch-Murnaghan EoS in every temperature range and gives better fitting in five cases out of the six in comparison to the Vinet EoS.

**4. Inner core densities**

Using the determined thermodynamic parameters for epsilon iron the inner core densities were calculated (Fig. 3). The new EoS predicts lower densities for the inner core than previous investigations (Saxena, 2003). Assuming 6000 K temperature in the inner core the EoS determined from all of the data gives 2.8-4.9% higher iron densities than the densities of the PREM model.

The separate clustering of the residuals of the Funamori et al. (1996) from the rest of the data indicates that the experiments contain a systematic error. Removing the Funamori's experiments from the data set the parameters of the New (6) EoS were determined. Calculating the densities from these new parameters the PREM inner core densities can be reproduced within 1% if 8000 K is assumed for the temperature of the inner core. The EoS parameters were also determined using the experiments of (Dubrovinsky et al. 1997; 1998-a-b; 2000 and Mao et al. 1987, 1990).

The calculated core densities from these experiments reproduce the PREM densities within one percent if 6000 K temperature is assumed for the inner core.

## 5. Conclusions

The author's new EoS was tested against experiments of epsilon iron. The pressure derivative of iron is not constant and the introduction of an exponential factor is required.

It is suggested that at pressures higher than the solid critical pressure the potential well becomes symmetrical and the temperature has no effect on the volume. The solid critical pressure for epsilon iron is around 400 GPa.

The parameters providing the best fit for epsilon iron are: $K_o = 116.31$ GPa, $V_o = 6.818$ cm$^3$, $\alpha_o = 2.823 \times 10^{-4}$ K$^{-1}$, a = 6.439, b = 0.7563 and c = $-7.119 \times 10^{-8}$ K$^{-1}$GPa$^{-1}$. The root-mean-square-deviations (RMSD) of the molar volume, pressure, and temperature are 0.021 cm$^3$, 2.0 GPa and 144.9 K respectively. These values are slightly higher than the uncertainties of the experiments.

Separating the experiments into 200 K temperature range the most widely used isothermal EoSs, Birch-Murnaghan, Vinet, and Roy & Roy equations were compared to the new EOS. Based on the RMSD and AIC values the new EOS is superior to the Roy-Roy and Birch-Murnaghan EoSs and equal or better than the Vinet EoS.

The new EoS predicts lower densities for the inner core than previous investigations. It is suggested that based on the currently available experiments it is too early to exclude the possibility of a pure iron-nickel core with no light elements.

**Acknowledgement**

This research was supported by Florida International University Dissertation Year Fellowship.

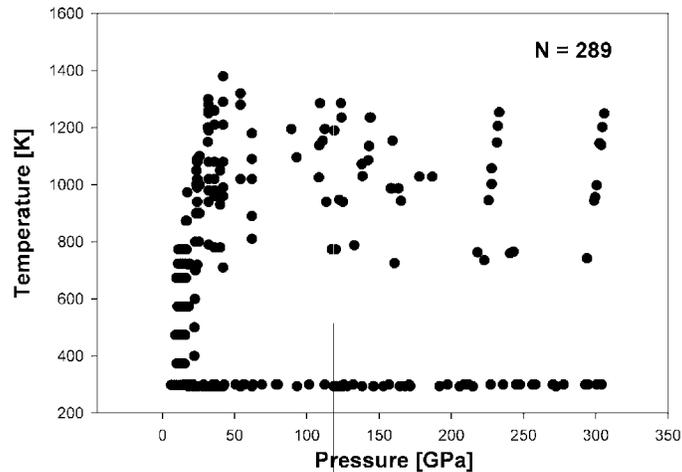

Fig. 1. Pressure-temperature range covered by the experiments.

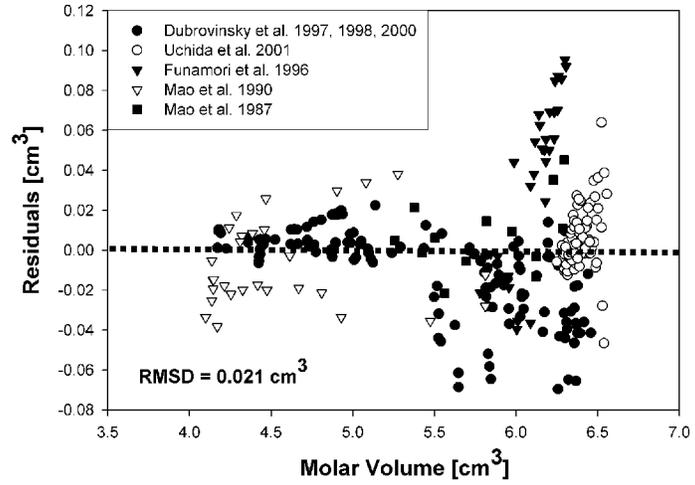
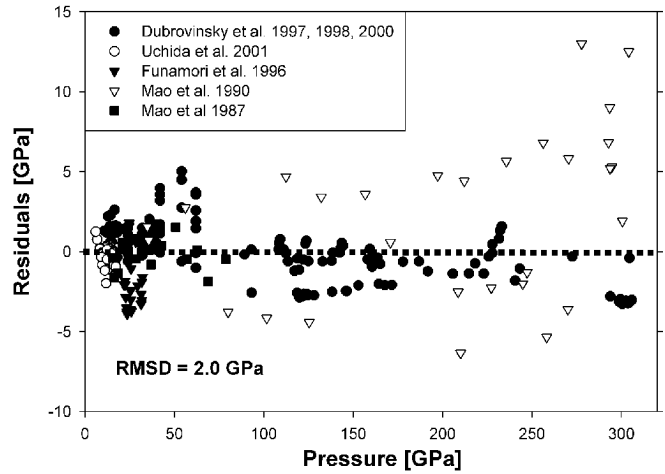
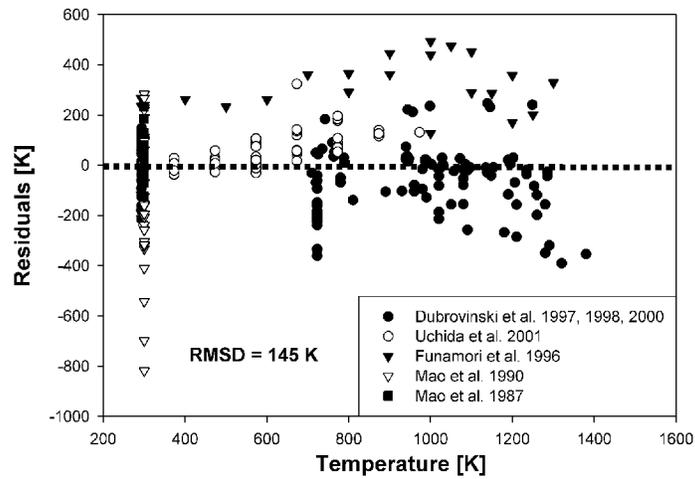

Fig. 2. Residuals plotted against (a) volume (b) pressure (c) temperature

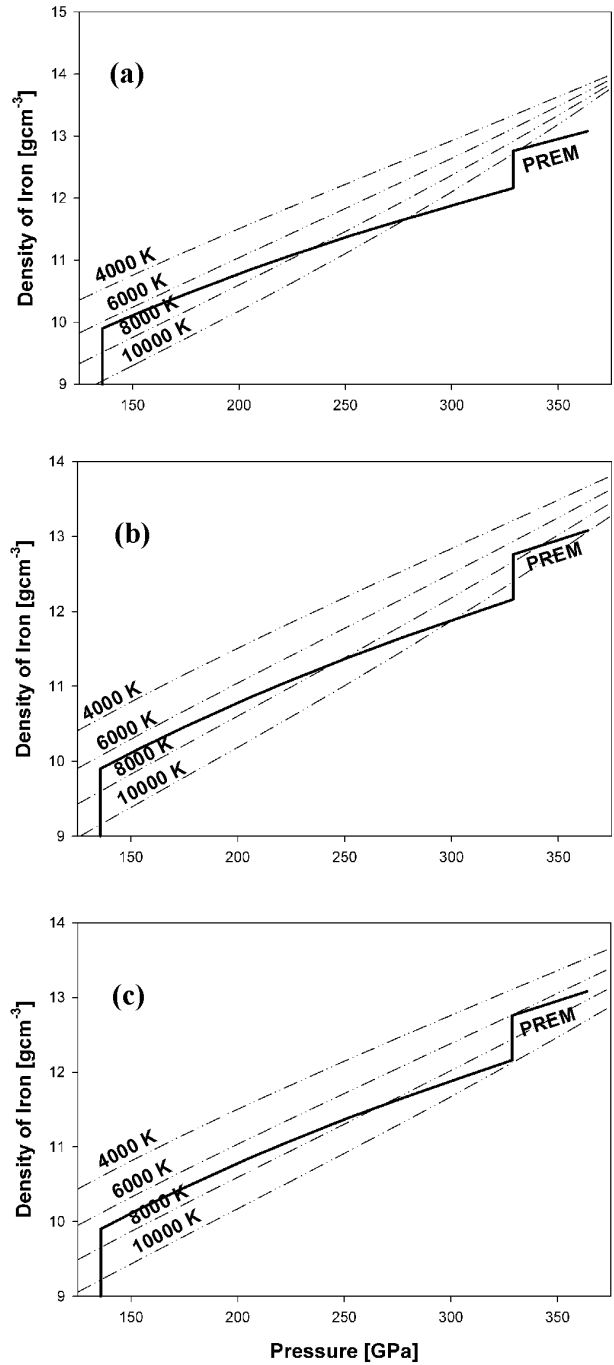

Fig. 3. The pure iron inner core densities were calculated from the EoS determined from (a) all the data (b) leaving out the data of Funamori et al. (1996) (c) from the data of Dubrovinsky et al. (1997, 1998, 2000) and Mao et al. (1987, 1990). The parameters of these EoSs are given in

Table 1. P-V-T and P-V fitting parameters and results. The values in the parentheses are fixed.

| Total 11-306 GPa 293 1573 K [N=289] | $K_o$ [GPa] | $V_o$ [cm³] | $\alpha_o$ [10⁻⁵K⁻¹] | a | b | c [10⁻⁸K⁻¹] | $K_{0T}$ [GPa] | $V_{0T}$ [cm³] | $K_{0T}'$ | $K_{0T}''$ | R | RMSD | AIC |
|---|---|---|---|---|---|---|---|---|---|---|---|---|---|
| V(p,T) New (5) | 228.72 | 6.736 | -58.76 | 1.340 | | 755.5 | | | | | 0.99703540 | 0.059 | -1623.1 |
| V(p,T) New (6) | 116.31 | 6.818 | 2.823 | 6.439 | 0.7563 | -7.119 | | | | | 0.99961944 | 0.021 | -2214.0 |
| V(p,T) New (6)*¹ | 110.39 | 6.839 | 2.545 | 7.110 | 0.7391 | -5.769 | | | | | 0.99969763 | 0.020 | -2034.5 |
| V(p,T) New (6)*² | 125.31 | 6.782 | 2.362 | 6.767 | 0.7453 | -4.746 | | | | | 0.99974405 | 0.017 | -1528.4 |
| p(V,T) New (6-0) | | | | | | | | | | | | 2.00 | 388.6 |
| T(p,V) New (6-0) | | | | | | | | | | | | 144.9 | 2866.0 |
| T = 300-500 K | | | | | | N = 124 | | | | | | | |
| V(p,T) New (6-0) | (116.3) | (6.818) | (2.823) | (6.439) | (0.756) | (-7.119) | | | | | | 0.018 | -999.2 |
| V(p,T) New (6-3) | 110.0 | (6.818) | (2.823) | 7.696 | 0.726 | (-7.119) | | | | | 0.99976827 | 0.017 | -1000.5 |
| V(p) (Roy-Roy (3) | | | | | | | 142.2 | (6.834) | 5.64 | -0.0467 | 0.99965315 | 0.021 | -950.5 |
| P(V,T) New (6-0) | | | | | | | | | | | | 2.647 | 241.5 |
| P(V,T) New (6-3-0) | | | | | | | | | | | | 2.180 | 193.3 |
| p(V) Birch (3) | | | | | | | 151.8 | 6.802 | 5.53 | | 0.99967659 | 2.317 | 214.4 |
| P(V) Vinet (3) | | | | | | | 138.1 | 6.834 | 6.20 | | 0.99969693 | 2.243 | 206.4 |
| T(p,V) New (6-0) | | | | | | | | | | | | 157.4 | 1254.6 |
| T(p,V) New (6-3-0) | | | | | | | | | | | | 134.6 | 1215.7 |
| 500-700 K | | | | | | N = 25 | | | | | | | |
| V(p,T) New (6-0) | | | | | | | | | | | | 0.019 | -197.0 |
| V(p,T) New (6-3) | 159.5 | (6.818) | (2.823) | 0.011 | 2.500 | (-7.119) | | | | | 0.99050764 | 0.012 | -215.7 |
| V(p) (Roy-Roy (3) | | | | | | | 115.0 | (7.012) | -0.015 | -0.0077 | 0.97541866 | 0.019 | -192.1 |
| P(V,T) New (6-0) | | | | | | | | | | | | 0.607 | -24.0 |
| P(V,T) New (6-3-0) | | | | | | | | | | | | 0.335 | -54.7 |
| p(V) Birch (3) | | | | | | | 119.6 | 6.997 | (5.53) | | 0.97692005 | 0.576 | -23.6 |
| P(V) Vinet (3) | | | | | | | 112.8 | 7.012 | (6.20) | | 0.97722507 | 0.572 | -23.9 |
| T(p,V) New (6-0) | | | | | | | | | | | | 104.2 | 223.0 |

| | | | | | | | | | | | | |
|---|---|---|---|---|---|---|---|---|---|---|---|---|
| T(p,V) New (6-3-0) | | | | | | | | | | | 62.4 | 206.7 |
| **700-900 K** | | | | | | N = 51 | | | | | | |
| V(p,T) New ( 6-0) | | | | | | | | | | | 0.028 | -365.3 |
| V(p,T) New ( 6-3) | 110.2 | (6.818) | (2.823) | 7.177 | 0.740 | (-7.119) | | | | 0.99923686 | 0.027 | -364.1 |
| V(p) (Roy-Roy (3) | | | | | | 178.9 | (6.923) | 0.331 | -0.0073 | 0.99323559 | 0.079 | -253.0 |
| P(V,T) New (6-0) | | | | | | | | | | | 1.141 | 13.5 |
| P(V,T) New (6-3-0) | | | | | | | | | | | 1.00 | 0.1 |
| p(V) Birch (3) | | | | | | 150.3 | 6.891 | 5.52 | | 0.99986659 | 1.175 | 22.5 |
| P(V) Vinet (3) | | | | | | 137.1 | 6.923 | 6.19 | | 0.99987218 | 1.150 | 20.3 |
| T(p,V) New (6-0) | | | | | | | | | | | 151.1 | 511.8 |
| T(p,V) New (6-3-0) | | | | | | | | | | | 142.1 | 505.6 |
| **900-1100 K** | | | | | | N = 46 | | | | | | |
| V(p,T) New ( 6-0) | | | | | | | | | | | 0.014 | -395.2 |
| V(p,T) New ( 6-3) | 125.5 | (6.818) | (2.823) | 5.027 | 0.798 | (-7.119) | | | | 0.99985376 | 0.012 | -402.8 |
| V(p) (Roy-Roy (3) | | | | | | 141.1 | (6.990) | 5.497 | -0.046 | 0.99978739 | 0.014 | -385.6 |
| P(V,T) New (6-0) | | | | | | | | | | | 1.208 | 17.4 |
| P(V,T) New (6-3-0) | | | | | | | | | | | 1.156 | 13.6 |
| p(V) Birch (3) | | | | | | 156.7 | 6.926 | 5.27 | | 0.99990935 | 1.12 | 16.3 |
| P(V) Vinet (3) | | | | | | 136.3 | 6.990 | 6.08 | | 0.99992615 | 1.01 | 6.9 |
| T(p,V) New (6-0) | | | | | | | | | | | 95.8 | 419.8 |
| T(p,V) New (6-3-0) | | | | | | | | | | | 79.3 | 102.4 |
| **1100-1300 K** | | | | | | N = 33 | | | | | | |
| V(p,T) New ( 6-0) | | | | | | | | | | | 0.024 | -247.5 |
| V(p,T) New ( 6-3) | 122.7 | (6.818) | (2.823) | 5.281 | 0.790 | (-7.119) | | | | 0.99950776 | 0.022 | -245.9 |
| V(p) (Roy-Roy (3) | | | | | | 146.7 | (7.011) | 5.35 | -0.0432 | 0.99945753 | 0.023 | -242.7 |
| P(V,T) New (6-0) | | | | | | | | | | | 1.75 | 36.9 |
| P(V,T) New (6-3-0) | | | | | | | | | | | 1.60 | 31.0 |
| p(V) Birch (3) | | | | | | 163.5 | 6.945 | 5.05 | | 0.99985426 | 1.57 | 35.9 |
| P(V) Vinet (3) | | | | | | 141.8 | 7.011 | 5.88 | | 0.99987007 | 1.49 | 32.2 |
| T(p,V) New (6-0) | | | | | | | | | | | 149.6 | 330.5 |
| T(p,V) New (6-3-0) | | | | | | | | | | | 132.0 | 322.3 |
| 1300-1500 K | | | | | | N = 8 | | | | | | |
| V(p,T) New (6-0) | | | | | | | | | | | 0.04 | -51.6 |
| V(p,T) New (6-3) | 195.2 | (6.818) | (2.823) | 0.0012 | 2.603 | (-7.119) | | | | 0.99795425 | 0.013 | -63.0 |
| V(p) (Roy-Roy (3) | | | | | | | 146.2 | (7.223) | 0.529 | -0.010 | 0.99615190 | 0.018 | -57.9 |
| P(V,T) New (6-0) | | | | | | | | | | | | 2.29 | 13.3 |
| P(V,T) New (6-3-0) | | | | | | | | | | | | 0.515 | -10.6 |
| p(V) Birch (3) | | | | | | | 134.8 | 7.223 | 3.90 | | 0.99706334 | 0.729 | 0.9 |
| P(V) Vinet (3) | | | | | | | 133.4 | 7.223 | 4.10 | | 0.99705558 | 0.730 | 1.0 |
| T(p,V) New (6-0) | | | | | | | | | | | | 220.4 | 86.3 |
| T(p,V) New (6-3-0) | | | | | | | | | | | | 72.6 | 68.6 |

[*1] The experiments of Funamori et al. 1996 are removed. (N = 260)
[*2] The experiments of Dubrovinsky et al. 1997, 1998, 2000 and Mao et al. 1987 and 1990 (N = 188)